\documentclass[twocolumn]{aa}
\usepackage{graphicx}
\begin{document}
   \title{The V-band luminosity function of galaxies in A2151}

   \author{R. S\' anchez-Janssen
	\inst{1}
   \and
	J. Iglesias-P\'aramo
	\inst{2}
   \and
	C. Mu\~noz-Tu\~n\'on
	\inst{1}
   \and
	J.A.L. Aguerri
	\inst{1}
   \and
        J.M. V\'\i lchez
	\inst{3}
          }
   \offprints{R. S\' anchez-Janssen\\
		\email{ruben@iac.es}}

   \institute{Instituto de Astrof\'\i sica de Canarias,
                 Calle V\'\i a L\'actea s/n, E-38205 La Laguna, Tenerife,
 Spain\\
                 \email{ruben@iac.es; cmt@iac.es; jalfonso@iac.es}
            \and
                Laboratoire d'Astrophysique de Marseille,
		BP8, Traverse du Siphon, F-13376 Marseille, France\\
                \email{jorge.iglesias@oamp.fr}
	    \and
	        Instituto de Astrof\'\i sica de Andaluc\'\i a,
		CSIC, Apdo. 3004, E-18080 Granada, Spain\\
		\email{jvm@iaa.es}
              }

   \date{Received ...; accepted ...}

   \abstract{We present a wide field $V$-band imaging survey of
     approximately 1 deg$^2$ ($\sim7.2~h^{-2}_{75}$ Mpc$^{2}$) in the
     direction of the nearby cluster of galaxies Abell 2151 (the
     Hercules Cluster). The data are used to construct the luminosity
     function (LF) down to $M_V \approx -14.85$, thus allowing us to
     study the dwarf galaxy population in A2151 for the first
     time. The obtained global LF is well described by a Schechter
     function with best-fit parameters $\alpha =
     -1.29^{+0.09}_{-0.08}$ and $M_V^* =  -21.41^{+0.44}_{-0.41}$. The
     radial dependence of the LF was investigated, with the faint-end slope
     tending to be slightly steeper in the outermost regions
     and farther away than the virial radius. Given the
     presence of significant substructure within the cluster, we
     also analysed the LFs in three different regions. We find that
     the dwarf to giant ratio increases from the northern to the
     southern subcluster, and from low to high local density
     environments, although these variations are marginally
     significant (less than 2$\sigma$).
 }


   \maketitle
%

\section{Introduction}

The luminosity function (LF) of galaxies (that is, the probability
density of galaxies in a certain population of a given luminosity)
is one of the key observable quantities for galaxy evolution
theories (Binggeli, Sandage \& Tammann 1988), although the link between the
data
and the underlying physics is still not easily understood (Benson et
al.\ 2003). Galaxies in high density environments, such as rich galaxy 
clusters,
can evolve rapidly (less than 1 Gyr) due to different physical
processes, e.g. harassment (Moore et al.\ 1996), ram-pressure stripping
(Gunn \& Gott 1972; Quilis, Moore \& Bower 2000), tidal effects and
mergers (Toomre \& Toomre 1972; Bekki, Couch \& Shioya 2001; Aguerri, Balcells 
\& Peletier
2001) or starvation (Bekki, Couch, \& Shioya 2002). The LF should
reflect this important role of the environment in regulating galaxy
evolution. Recent results from large surveys (De Propris et al.\ 2003)
point towards the composite LF of cluster galaxies having a
characteristic magnitude that is approximately 0.3 mag brighter and a
faint-end slope (directly related to the ratio of dwarf to giant
 galaxies, D/G) that is
approximately 0.1 steeper than those found for field galaxies. As 
hierarchical galaxy formation
theories predict the formation of large numbers of low mass halos, the 
latter observational
evidence may indicate the existence of some mechanism that eliminates at 
least the visible component
of galaxies in low density regions (Tully et al.\ 2002).

A2151 (the Hercules cluster), together with A2147 and A2152, forms part of
 the
Hercules Supercluster, one of the largest and most massive 
structures
in the local Universe (Chincarini, Thompson \& Rood 1981; Barmby \& 
Huchra 1998). The kinematical analysis of
Barmby \& Huchra (1998) reveals that A2151 and A2147 are probably bound 
to each
other, and that the supercluster as a whole may also be bound. A2151 
($z= 0.0367$) is an irregular and spiral-rich cluster ($\sim$ 50\%, 
Giovanelli \&
Haynes 1985). There is strong evidence (from optical and X-ray
studies) suggesting that the cluster is still in the process of
collapsing: the lack of hydrogen deficiency in the spiral population
(Giovanelli \& Haynes 1985; Dickey 1997), the bumpy distribution of the 
hot
intracluster gas and its low X-ray flux (Magri et al.\ 1988; Huang \& 
Sarazin 1996), and
the presence of at least three distinct subclusters (Bird, Davis \&
Beers 1995, BDB hereafter). All this evidence indicates that  A2151 is
 a young and
relatively unevolved cluster, thus making it an excellent target for 
studying the LF and establishing comparisons with more evolved
systems. Throughout this article we assume a value of $H_{0} = 75$ km 
s$^{-1}$ Mpc$^{-1}$,
which puts A2151 at a distance of 147 Mpc.

The aim of this paper is to study the LF of A2151 down to the dwarf
regime,thus extending the previous work of Lugger (1986), who computed the $R$-band 
LF of the cluster over a wide area (4.6 $\times$ 4.6 Mpc$^2$) but only down to
$M_R = -19.8$. The paper is arranged as follows: in Section 2 we describe the 
observations
and data reduction procedures. Section 3 gives the details of the source 
extraction and  star/galaxy
separation, and the estimate of the limiting magnitudes of the sample. 
In Section 4 we present the
LF in A2151 and analyse possible variations of its parameters within 
location in the cluster. Finally, Section 5 discusses the results.

\begin{table*}
      \caption[]{Log of the observations. Seeing values are averaged
from the four chips in each of the fields.}
         \label{log}
     $$
         \begin{tabular}{lccccc}
            \hline
            \noalign{\smallskip}
Field & R.A.$^{a}$ & Dec.$^{a}$ & Date & $t_{exp}$ & Seeing \\
      & (J2000) & (J2000) &      & (sec)     & (arcsec) \\
            \noalign{\smallskip}
            \hline
            \noalign{\smallskip}
Ce1  & 16:05:36.00 & $+$17:43:59.97 & 1999 June 06  & $2 \times 900$  & 
1.2 \\
Ce2  & 16:05:23.99 & $+$17:27:59.97 & 1999 June 06  & $2 \times 900$  & 
1.5 \\
\#2  & 16:06:12.56 & $+$18:10:56.15 & 2001 April 17  & $2 \times 900$  & 
1.1 \\
\#4a & 16:02:00.00 & $+$17:55:43.11 &  2002 May 19  & $ 2 \times 900$  & 1.7
 \\ 
\#4b & 16:04:00.00 & $+$17:50:42.97 & 2002 May 19  & $2 \times 900$  & 1.6
 \\
Back & 15:39:13.99 & $-$00:13:01.07 & 2002 July 11  & $2 \times 1000$ & 1.7 \\
         \noalign{\smallskip}
            \hline
         \end{tabular}
     $$
\\
$^{a}$ Coordinates correspond to the centre of the WFC. See http://www.ing.iac.es/Astronomy/instruments/wfc/index.html for a detailed description of the spatial arrangement of the mosaic.
\end{table*}


\section{Observations and data reduction}

Observations were carried out with the Wide Field Camera at the 2.5 m
 Isaac Newton Telescope (INT), located at the Observatorio del Roque de 
 los Muchachos during four different runs in the period  1999--2002.
 The WFC consists of four thinned AR-coated EEV 4k $\times$ 2k CCDs and a 
 fifth one acting as an autoguider. The pixel scale is 0.33 arcsec 
 pixel$^{-1}$, giving a total field of view of about 34 $\times$ 34 
 arcmin$^{2}$. A 
square area of about 11 $\times$ 11 arcmin$^{2}$ is lost at the top right 
corner of the field due to the particular arrangement of the detectors. 
The top left corner of Detector \#3 is also lost because of filter 
vignetting.

We imaged a mosaic of five overlapping pointings covering a total 
area of about 1.05 deg$^2$ ($\sim7.2~h^{-2}_{75}$ Mpc$^{2}$) in the 
direction of the Abell~2151 cluster. The total exposure time for each
field was 2 $\times$ 900 s using the Harris $V$ broad-band filter.
A control field outside the Abell~2151 cluster was also observed in order
 to get the background galaxy counts with the same instrumental 
 configuration as the cluster counts. Two 1000~s exposures were taken in the 
 direction of the Landolt~107 field. The average seeing of the fields ranged 
 between 1.1 and 1.7 arcsec, with a maximum value of 2 arcsec in one of 
 the
 CCDs. Table~\ref{log} shows a detailed log of the observations.

Data reduction consisting of bias subtraction and flat-fielding was 
carried out in the usual way using the IRAF reduction package.\footnote{IRAF 
is distributed by the National Optical Astronomy Observatories, 
operated by the Association of Universities for Research in Astronomy, 
Inc., under cooperative agreement with the National Science Foundation.} 
The images corresponding to the Abell~2151 field were calibrated using 
aperture magnitudes of bright galaxies taken from Takamiya, Kron \& Kron 
(1995) since no standard stars were available for the photometric 
calibration (see Table~1). The differences between our calibrated magnitudes
and those of TKK95 for galaxies in our central field (computed within an 
aperture of 17.5~arcsec radius) never exceeded 0.05~mag. Photometric zero 
points for the remaining fields were calculated by cross-correlating 
aperture
 magnitudes of common objects in the overlapping chips, for which we 
 obtained a maximum dispersion of 0.18~mag at the faintest magnitudes. On the 
other hand, the background field was accurately calibrated since several 
standards from the Landolt~107 field (Landolt 1992) show up in the 
different frames. All the $V$ magnitudes were corrected for
 Galactic extinction 
using the Schlegel, Finkbeiner \& Davis (1998) dust maps and the 
Cardelli, Clayton \& Mathis (1989) extinction curve.


\section{Data Analysis}

\subsection{Source extraction}

Identification and extraction of sources was carried out using the 
automated code SExtractor
(for specific details see Bertin \& Arnouts 1996).
Objects are identified as every detection having a certain minimum area 
and number counts above a limiting threshold 
taking the sky local background as a reference. These limiting sizes and 
fluxes were set to 36 pixels~\footnote{this corresponds to an
apparent size of 2 arcsec, which is the maximum seeing disc size 
in our frames.} and twice the standard deviation
of the sky counts, respectively. More precisely, sources were extracted 
using a background-weighted method, so that in every
case the threshold takes a local value. This avoids a lot of spurious detections in
the surroundings of bright galaxy halos and saturated stars, where the 
local threshold is significantly increased. Careful visual inspection  of 
the frames was carried out and all   definitely spurious entries were
 removed from the catalogues, such as H\ {\sc ii} regions of spiral 
 galaxies, which fulfil all the above-stated requirements.

The separation between stars and galaxies was performed on the basis of the 
SExtractor stellarity index (S/G hereafter). The code assigns a numerical value
 between 0 and 1 to all detections according to the following 
 criteria: star-like objects have values close to 1 while galaxy-like objects 
 are closer to 0. This S/G classifier behaves correctly for bright 
 objects but breaks down at the lower magnitude end (especially in bad seeing 
conditions), where a correct classification remains difficult.

 In order to test the reliability of the index we simulated artificial
 \emph{isolated}~\footnote{we stress this fact because the S/G
 classifier has more problems when classifying stars overlapping with
 other objects or lying in galactic halos.} stars and recovered them
 using SExtractor and the same selection criteria as for the
 Hercules fields. The stars were generated with the
 artdata.mkobjects package in IRAF and then added into frames
 reproducing the same sky counts and S/N ratio as the scientific ones. The PSF
 profiles were assumed to be Gaussian~\footnote{we also tried  with a more realistic
 Moffat function, but differences in limiting magnitude and S/G values
 were negligible.} and FWHMs spanned the values of our seeing
 conditions. A total number of 750 stars were simulated in the magnitude 
range
17 $\le m_V \le$ 23 for two different seeing conditions (1.2 and
 2 arcsec FWHMs).

  \begin{figure}[t]
   \centering
   \includegraphics[width=9cm,height=7cm]{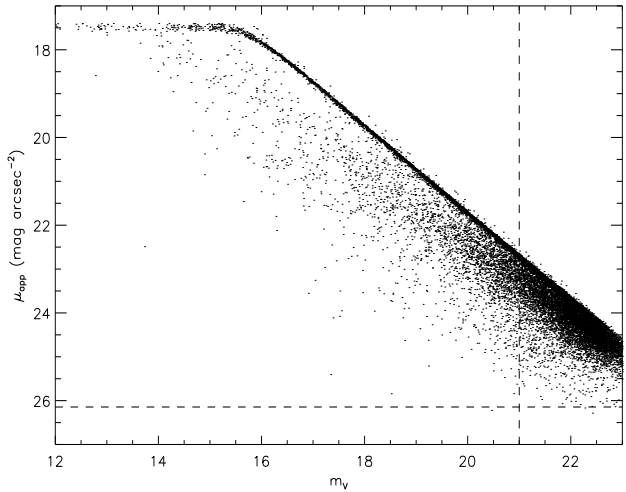}
      \caption{The aperture central surface brightness of all our detected
	objects as a function of apparent magnitude. Note the different location
	in the plot of stars and extended objects. The horizontal
	dashed line corresponds to the 2$\sigma$ isophotal detection
	limit used in SExtractor, which cuts the objects' distribution at $V \sim 21$
	(vertical dashed line), thus setting the limiting magnitude of the sample 
	(see text for details).
	}
      \label{mu}
  \end{figure}

 The classification of star-like objects is
 easily done for good seeing frames (S/G $\ge 0.95$), whilst the 
 S/G values span a wider range in worse
 conditions (S/G $\ge 0.85$ for simulated point-like sources
 brighter than $m_{V}=22.0$). Photometry in all the cases was 
 almost perfect, with
 differences between input and recovered magnitudes of less than 0.05
 mag even for the faintest stars. As will be shown in Sect.\ 3.2.1, 
 our limiting magnitude for cluster galaxies is $m_V \sim$ 21. Thus, we
 cautiously set a value  of S/G $\ge$  0.85 for the star-like
 objects in all the frames, which corresponds to the S/G value of
 simulated stars one magnitude fainter than our limiting magnitude. All 
 the detections having 0.8 $\le$  S/G $<$ 0.85 were carefully inspected
by means of IMEXAM, along with bright saturated stars that tend to have lower 
S/G values. Those detections with obvious stellar origin  were removed 
from the catalogues. After this exercise we expect
the lowest contamination by stars possible, although perfect 
\emph{cleaning} is almost unachievable
for the faintest magnitudes. There is a further concern about losing 
compact dwarf galaxies as
 so-called ultra compact dwarfs (UCDs; Drinkwater et al.\ 2000), which 
can assume a stellar appearance in ground-based images. However, these 
galaxies typically
have intrinsic magnitudes much fainter ($M_B \sim -11$) than our limiting
 magnitude and the dwarfs we
are able to detect have low probability of being misclassified (see Sect.~3.2.1).

\subsection{Photometry}

\begin{figure}
   \centering
   \includegraphics[width=9cm]{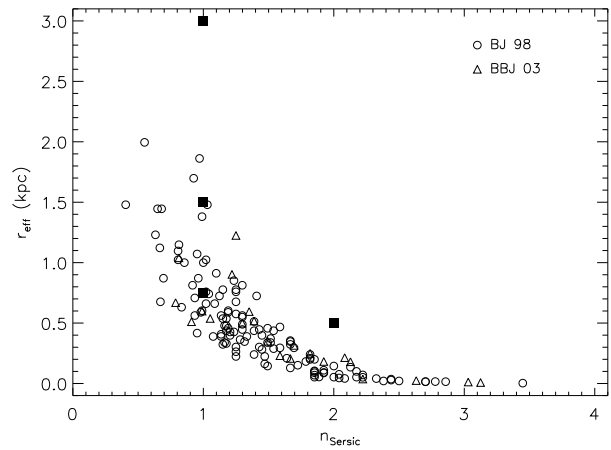}
      \caption{Structural parameters of dE and dSph galaxies in the Virgo
 Cluster. The
	data are taken from Binggeli \& Jerjen (1998) (\emph{open circles}) and 
Barazza,
	Binggeli \& Jerjen (2003) (\emph{open triangles}) while  filled squares 
correspond
	to our simulated galaxies. The object at the top left represents a de 
	Blok LSB
	galaxy with an exponential disc profile and a scale-length of 3 kpc.
	}
      \label{Sersicparams}
   \end{figure}

\begin{figure*}
   \centering
   \includegraphics[width=14cm,height=11cm]{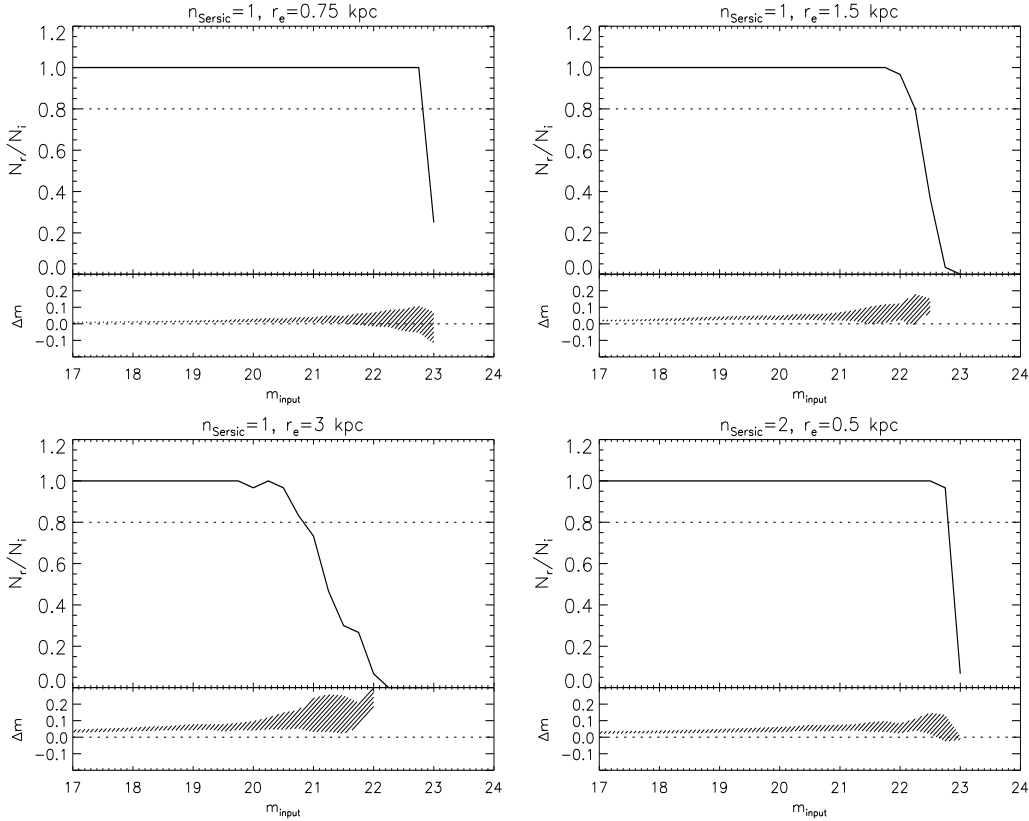}
      \caption{Detection efficiency (\emph{solid line}) for each of our 
        simulated galaxy models.
	From left to right and top to bottom: n  =1 with $r_e$ = 0.75, 1.5, and
        3 kpc, and the n = 2 with
	$r_e$ =  0.5 kpc model. The shaded regions in the lower panels 
	indicate differences between
	recovered and input magnitudes, which never exceed 0.3 mag.
	}
      \label{simsdetection}
   \end{figure*}

\subsubsection{Sample completeness}

One of the key problems in the study of LFs is  knowledge of
detection limits and completeness of the data. In order to make this
estimation we followed two different approaches.

First,  for all our detected objects we computed the mean central surface 
brightness ($\mu_{app}$) measured in
a circular aperture of area equal to the minimum detection area used in
SExtractor. Figure~\ref{mu} shows $\mu_{app}$ as a function of apparent
magnitude for all the detected sources (stars and  galaxies). Stars occupy 
the upper diagonal locus, with saturated objects defining the flat region of 
the distribution. Galaxies are located in the lower-most broader region of the plot.

The different locations of point-like and extended objects is evident at bright
magnitudes ($m_{V} \le 18$), while the effects of seeing make
 this distinction at fainter magnitudes difficult. We have computed the
 2$\sigma$ isophotal detection limit used in SExtractor, which
 corresponds to $\mu =26.15$ mag arcsec$^{-2}$ (horizontal dashed line
 in Fig.~\ref{mu}). Given the
 distribution of the objects in Fig.~\ref{mu}, $m_{V} \sim 21$ appears to
 be the appropriate limiting magnitude of our sample, as we do not
 lose objects because of their low surface brightness.

Secondly, we simulated artificial galaxies and, as we did with stars,
placed them in frames with conditions (seeing, S/N) identical to the
scientific ones, in order to recover them with SExtractor. We generated 
galaxies with two different morphological types and
different intrinsic half-light radii (one exponential profile with
$r_{e}$ = 0.75, 1.5 and 3 kpc, and one S\' ersic profile (S\'ersic 1968) 
with n = 2 and
$r_{e}$ =  0.5 kpc) in the magnitude range 17 $\le m_V \le$ 23 ($-$18.85 
$\le$
$M_V$ $\le$ $-$12.85) and with three different axial ratios (0.3, 0.6, and
0.9). A total number of 1500 galaxies were generated for each of the 
morphologies in order to get enough statistics.
The choice of these models was motivated by the quantitative morphological
studies of the populations that are supposed to dominate the faint-end of
 the LF in
nearby clusters (dE and dSph mainly), which show that these galaxies
lie in a well defined region on a $r_{e}$ vs.\ n
plot. Figure~\ref{Sersicparams} shows structural parameters of dE and
dSph galaxies in the Virgo Cluster taken from Binggeli \& Jerjen
(1998) and Barazza, Binggeli \& Jerjen (2003) (open circles and open
triangles, respectively), while filled squares correspond to our
artificial galaxies. Objects with larger effective radii are known to 
present
fainter surface brightness, which renders their detection difficult. For this
reason we modelled one object with n = 2 and $r_e$ slightly larger than those
observed and another galaxy with n = 1 and $r_e$ =  3 kpc; the 
latter are typical
parameters of the low surface brightness (LSB) galaxies studied by de Blok et
 al.\ (1995).
Note (see Fig.\ 2) the apparent lack of this kind of galaxy in this 
Virgo sample. LSB galaxies are
known to be the most unstable systems during cluster evolution,
suffering from rapid encounters and strong tidal shocks (harassment,
see Moore et al.\ 1999), therefore rare in relaxed clusters.
 However, as mentioned, the Hercules cluster is still
in process of collapse and harassment might have produced little
effect on such a kind of LSB. On the other hand Andreon \& Cuillandre 
(2002)
show that a large population of low surface brightness galaxies 
dominate the
LF in Coma, representing the main contribution at faint magnitudes.
As it is not obvious which kind of galaxy is the predominant one in the 
cluster environment,
we cautiously set the limiting magnitude of the lowest surface brightness
 model as our sample limit.

The results of our simulations are summarized in
Fig.~\ref{simsdetection}, where we plot the fraction of recovered
galaxies ($N_r/N_i$) vs.\ input apparent magnitude (as given by
MAG\_BEST) for each galaxy type (solid lines); shaded regions
correspond to recovered magnitude differences, which are always below
0.3 mag. We find that the detection and estimate of magnitudes are less 
accurate for galaxies with larger effective radii than smaller ones.
Within each of the morphological types, edge-on galaxies
are more easily detected than face-on ones, but the differences are small
enough that we could focus our analysis on averaged numbers. All these
results were expected, for it is widely known that detection
limits strongly depend on the interplay between magnitudes, effective
radius and inclination. The recovered S/G parameters for the simulated
sample were always smaller than 0.75. From Fig.~\ref{simsdetection} it
is apparent that detection efficiencies strongly depend on the
structural parameters of galaxies.

Our limiting magnitude, as derived from the $\mu_{\rm app}-m_{V}$ 
relation, is
$m_V \sim$  21 ($M_V \sim -14.85$ at A2151 distance). From the 
simulations, this value ensures that our sample
is at least  80\% complete, with the worst case that of LSB
galaxies.  In order to construct the LF of A2151, only objects brighter
 than this limiting magnitude were considered.

\subsection{The control field}

\begin{figure}
   \centering
   \includegraphics[width=9cm]{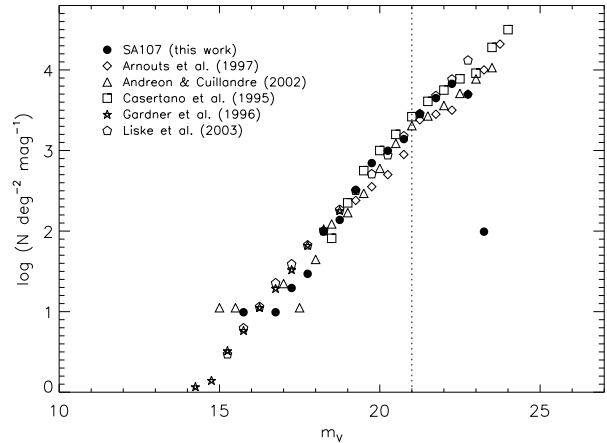}
      \caption{Background galaxies number counts in the SA107 field 
      (\emph{filled circles}), compared
	with other sources in the literature. The limiting magnitude of LSB 
	galaxies in
	the A2151 sample is marked with the dotted vertical line.
	}
      \label{back_counts}
   \end{figure}

The most reliable method for computing deep LFs of cluster galaxies in
our type of survey is by means of a statistical subtraction of
background/foreground number counts from a control field, since the
spectroscopic information is usually limited to the brightest galaxies.
\footnote{Redshift data in the Supercluster are only complete down to 15.1 mag,
see Barmby \& Huchra (1998).} It should be noted that the selection of the
control field is critical, for the final results derived from the LF
strongly depend on this choice. The selected field must fulfil the
requirement that the area is large enough so the cosmic variance and
number count statistics, the main error sources in LF
determination, are minimized. We imaged the SA107 ($\alpha =  15^h
39^m 12^s$, $\delta = -00^{\mathrm{o}} 19' 60''$) flanking field for
these purposes, where the presence of photometric standards provided
us with high photometric precision for the background counts. In
    order to ensure reliable subtraction, the background field
    should contain a population of objects similar to that of the cluster. We
    checked this by constructing a plot like the one in Fig.~\ref{mu},
with the result that the population of objects in the background and in our
cluster fields were similar. The number counts in this field are 
plotted in Fig.~\ref{back_counts}, together with other $V$-band counts taken from
 different sources in the literature.\footnote{The data from the MGC of 
 Liske et al.\ (2003) were transformed from the $B$-band using their average $B-V$
 =  0.94.} The fair agreement between these counts, each
 taken with different instrumentation under distinct conditions, makes us
 confident that the resulting LF derived by subtracting our control field counts 
is significant, with the advantage of having been
 obtained under the same instrumental set-up as the cluster counts.

\begin{table*}
      \caption[]{Best-fit values for the radial LF of A2151. The 
      quantities denoted by $k$ are derived from an exponential fit to the 
      faint-end
 of each LF. The columns correspond to the values derived for the three 
 Schechter parameters, those obtained for the exponential parameters, the 
 number of galaxies ($N_{\rm gal}$) used in the computation of each LF and its
D/G ratio.
}
         \label{radialparams}
     $$
         \begin{tabular}{lcccccccc}
            \hline
            \noalign{\smallskip}
Region & $\alpha$  & M* & $\phi^{*}$ & $\chi_{\nu}^{2}$ & $\alpha_{k}$ & 
$\chi_{k}^{2}$ & N$_{gal}$ & D/G\\
            \noalign{\smallskip}
            \hline
            \noalign{\smallskip}
$r~\le~r_{c}$                    & $-1.22 \pm 0.12$ & $-22.29 \pm 2.09$ &
 $149 \pm 149$ & 0.77  & $-1.18 \pm 0.06$ & 1.25 & 735 & 3.54 $\pm$ 0.76\\
            \noalign{\smallskip}
$r_{c}~<~r~\le~(r_{c}+r_{v}$     & $-1.26 \pm 0.09$ & $-21.82 \pm 0.49$ &
 $42 \pm 18$ & 0.53  & $-1.26 \pm 0.11$ & 0.96 & 5575 & 3.91 $\pm$ 0.78\\

            \noalign{\smallskip}
$(r_{c}+r_{v})/2~<~r~\le~r_{v}$  & $-1.27 \pm 0.12$ & $-20.63 \pm 0.50$ &
 $65 \pm 33$ & 0.59  & $-1.21 \pm 0.05$ & 0.29 & 4248 & 7.05 $\pm$ 1.58\\

            \noalign{\smallskip}
$r~>~r_{v}$                      & $-1.47 \pm 0.26$ & $-21.63 \pm 4.40$ &
 $13 \pm 39$ & 0.76  & $-1.32 \pm 0.13$ & 0.41 & 2466 & 12.09 $\pm$ 5.11\\
            \noalign{\smallskip}
            \hline
         \end{tabular}
     $$
   \end{table*}

\begin{table*}
      \caption[]{Best-fit values for the LF of A2151 and the three 
      studied subclusters.}
         \label{subparams}
     $$
         \begin{tabular}{lcccccccc}
            \hline
            \noalign{\smallskip}
Region & $\alpha$  & M* & $\phi^{*}$ & $\chi_{\nu}^{2}$ & $\alpha_{k}$ & 
$\chi_{k}^{2}$ & N$_{gal}$ & D/G\\
            \noalign{\smallskip}
            \hline
            \noalign{\smallskip}
Global & $-1.29^{+0.09}_{-0.08}$ & $-21.41^{+0.44}_{-0.41}$ & $47 \pm 14$
 & 0.78  & $-1.24 \pm 0.09$ & 0.54 & 13024 & 5.13 $\pm$ 0.82\\
            \noalign{\smallskip}
A2151C & $-1.26 \pm 0.08$ & $-21.78 \pm 0.51$ & $70 \pm 29$ & 0.66  & 
$-1.17 \pm 0.04$ & 1.54 & 3380 & 3.95 $\pm$ 0.69\\
            \noalign{\smallskip}
A2151N & $-1.16 \pm 0.17$ & $-22.37 \pm 2.23$ & $54 \pm 55$ & 0.21  & 
$-1.15 \pm 0.08$ & 0.26 & 1880 & 2.63 $\pm$ 0.75\\
            \noalign{\smallskip}
A2151S & $-1.59 \pm 0.10$ & $-23.05 \pm 2.01$ & $8 \pm 11$ & 0.43  
& $-1.65 \pm 0.19$ & 0.55 & 687 & 9.38 $\pm$ 3.21\\
            \noalign{\smallskip}
            \hline
         \end{tabular}
     $$
   \end{table*}

\section{The Luminosity Function}

The luminosity function of galaxies in A2151 was computed as the 
statistical difference between the counts in the cluster and control field 
samples. Once again we stress that final results obtained for the LF 
will strongly depend on the galaxy counts used for foreground/background 
decontamination. Thus, in order to minimize this influence, we used 
large bin widths (1 mag) compared to photometric uncertainties (see Sect.\
2 and Sect.\ 3.2.1), and checked that galaxy counts in the SA107 field 
remained stable from chip to chip.

\begin{figure}
   \centering
   \includegraphics[width=9cm]{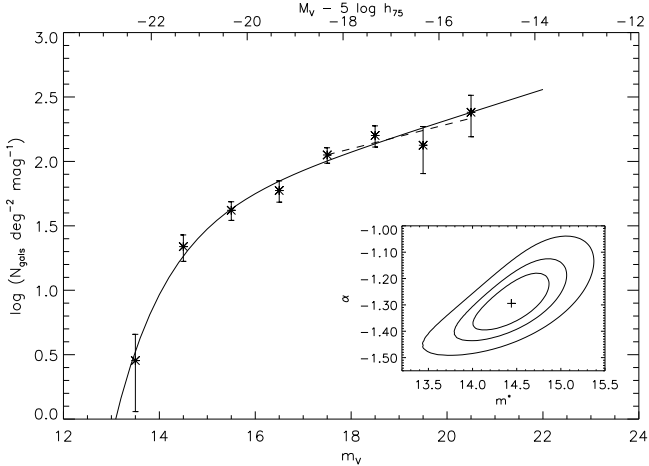}
      \caption{Total luminosity function of A2151 down to {$m_{V}$} = 21 
      (asterisks), with best-fit Schechter function
	(\emph{solid line}) also overplotted. The dashed line corresponds to the
 best fit to an exponential
	function for the faint-end of the LF ($-18.85 < M_{V}  < -14.85$).
	 Error bars include
	both Poissonian and non-Poissonian fluctuations of counts. The inset 
	shows the 1, 2 and 3$\sigma$
	contour levels for the best-fit Schechter parameters (indicated by the
cross).}
      \label{A2151LF}
   \end{figure}


Following Huang et al.\ (1997), errors from both Poissonian and
non-Poissonian fluctuations of counts have been taken into account.
Inclusion of this cosmic variance points to the fact that background
counts in the cluster line of sight are not \emph{exactly} those in
our control field. This error source affects both the cluster and
control field samples (Andreon \& Cuillandre 2002). Our final error bars 
include this latter term, along with both the contribution of photometric 
error of the zero point and the Poissonian uncertainties of both samples. 
For simplicity, all these sources were added in quadrature despite of 
their different nature.

After background galaxy counts subtraction, the data were fitted to a 
Schechter function (Schechter 1976):

\begin{equation}
\label{eq:schechter}
\phi(m_{V}) = \phi^{\ast} \times [10^{0.4(m^{\ast} - m_{V})}]^{\alpha +1}
e^{-10^{0.4(m^{\ast} - m_{V})}}.
\end{equation}

{\flushleft{Our fitting algorithm minimizes the $\chi^{2}$ by taking errors 
into account and assigns a statistical weight to each of the points 
(1/$\sigma_{i}^{2}$, where $\sigma_{i}$ corresponds to its error term).}}

In Fig.~\ref{A2151LF} we plot our LF for the total area covered in 
this study down to $m_V =$  21 mag. We find that the Schechter function is
a good representation of the LF in A2151 ($\chi_{\nu}^{2} =$  0.69) 
with best-fit parameters $\alpha = -1.29^{+0.09}_{-0.08}$~, M$_{V}^{*}  =
-21.41^{+0.44}_{-0.41} $ and $\phi^{*} = 47 \pm 14$.

In order to make a better estimate of the faint-end slope of the LF, we 
fitted our data to an exponential function of the type $\sim 10^{km}$ 
where $m$ are magnitudes and $k$ is related to the $\alpha$ parameter of the
Schechter function by the following expression:

\begin{equation}
\label{eq:alphapot}
\alpha = -(k/0.4 + 1).
\end{equation}

{\flushleft{We derive a best-fit value of $k = 0.095 \pm 0.036$ ($\chi_{\nu}^{2} =$ 
 0.54) in the magnitude range $-18.85 < M_{V}  < -14.85$, which
 corresponds to $\alpha_{k} = -1.24 \pm 0.09$ and is thus  consistent with the 
value derived from the Schechter function.}}

Lugger (1986) computes the $R$-band LF of A2151 in a region 4.6 Mpc a
side, but only down to $M_{R} \sim -19.8$, which is approximately five
magnitudes brighter than our limit. She derived a value of $\alpha =
-1.26 \pm 0.13$, similar to ours despite of other big differences
between both studies: photographic plates vs.\ CCD images, different
surveyed areas and limiting magnitudes, different bands.

It is well known that the $\alpha$ and M$^*$ parameters of the Schechter 
function show a high degree of
correlation (Schechter 1976), so that their values and errors are coupled.
In order to avoid this effect, we
computed the dwarf to giant ratio (D/G) in A2151 from the LF itself. 
Dwarf galaxies were defined as those
having magnitudes $-18 < M_B < -14.15$, while giants were those 
brighter than $M_B = -18$. The transformation
to $B$-band magnitudes was carried out using the average $B-V = 0.7$ for 
galaxies in nearby clusters in the GOLDMine
database (Gavazzi et al.\ 2003). We obtained a value of D/G = 5.13 $\pm$
 0.82 over the whole surveyed area. Errors
were computed using a Monte Carlo method, generating 100 LFs uniformly 
distributed within the
1$\sigma$ uncertainty region, and then computing the D/G for all of them. 
The final error was taken as the dispersion
of this distribution.

In the following subsections we investigate the dependence of the LF on 
environment within the cluster.

\subsection{Radial Analysis of the LF}

The dependence of the LF on the cluster region has been studied by
several authors, mainly for nearby clusters (but see Andreon 2001 and
Pracy et al.\ 2004 for this kind of study at intermediate redshift 
clusters). The pioneering work of Lugger (1989) shows that the faint-end of 
the LF tends to be flatter in high-density regions than in low-density 
regions. More recently, various authors have claimed the faint-end slope is 
steeper as we move outwards within the cluster than in the inner regions (e.g.\ 
 Sabatini et al.\ 2003 for Virgo, Beijersbergen et al.\ 2002 and 
 Iglesias-P\'aramo et al.\ 2003 for Coma). Independent studies (Aguerri et al.\
  2004) 
have shown that the core of the Coma Cluster is indeed dwarf-depleted 
when compared with the outer regions, giving rise to the observed steepening.

We therefore wanted to explore this trend in A2151. For this purpose, we 
divided our area in four different annuli from inside the core radius, 
$r_c$,  to outside the virial radius, $r_v$ (0.24$h^{-1}$ Mpc and 1.5$h^{-1}$ 
Mpc respectively, see Girardi et al.\ 1998) and computed their LFs. 
As the X-ray emission from the cluster shows a very irregular and bumpy 
structure, the cluster centre was assumed coincident with the optical 
centre of mass ($\alpha = 16^h 05^m 26^s$, $\delta = 17^{\mathrm{o}} 47' 
50''$) derived by Tarenghi et al.\ (1980). A summary of the results is 
listed in Table~\ref{radialparams}, where it can be seen that there is a 
tendency of increasing faint-end slope as we move outwards in the cluster. 
However, the large  uncertainties in the $\alpha$ parameter do not allow 
us to claim this trend with significance. It is worth noticing that the 
$M^*$ parameter is highly unconstrained in both the innermost and outermost
 annuli, showing that these LFs are better described with an exponential 
function rather than with a Schechter one.  Because of the limited 
statistics (small area covered), the number counts of background objects become
 almost equal to  A2151 counts in the most external region, so that the 
 LFs in the outermost annuli suffer from small number statistics, leading 
 to large errors in  slope determination. These cases are where
 non-parametrical estimation of the D/G ratio provides a more robust 
 study of luminosity segregation.

\begin{figure}
   \centering
   \includegraphics[width=9cm]{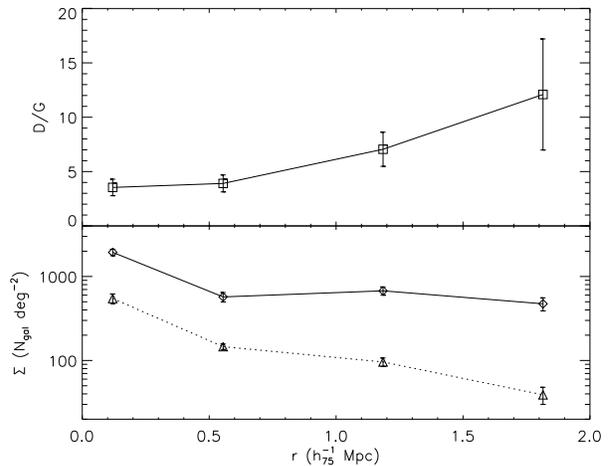}
      \caption{\emph{Top}: dwarf to giant ratio (D/G) vs.\ projected 
      clustercentric distance. Errors were computed by a
	Monte Carlo method.
	\emph{Bottom}: Radial surface density of dwarf (\emph{solid + diamonds})
 and giant galaxies (\emph{dotted + triangles}).
	Note that the increasing D/G as we move outwards in the cluster is due 
	to a more rapid decrease in the giant population,
	while dwarf galaxies show a more extended distribution.
	}
      \label{D2G}
   \end{figure}

We computed the radial dependence of the D/G ratio in these four annuli 
as in Section 4. The result is plotted
in the upper panel of Fig.~\ref{D2G}, where the aforementioned 
trend of increasing D/G ratio as we move outwards within the cluster is 
apparent. In order to determine the origin of this 
increase, we analysed the radial density profiles of the dwarf
and giant populations within A2151. The lower panel of Fig.~\ref{D2G} 
shows that the variation is mainly due to a
more rapid radial decrease in the giant population (dotted line + 
triangles), while dwarf galaxies (solid + diamonds)
present a more extended distribution (flatter profile). It is also clear 
from Fig.~\ref{D2G} that in the core of A2151, dwarf
galaxies follow the same behaviour as the giant population so that it 
appears not to be dwarf-depleted.
Lugger (1989) studies the radial dependence of the LF in A2151 by 
computing it in different annuli out to 1.2 Mpc from the cluster
centre. She finds that all LFs were consistent with that of the global 
cluster, so that no significant variation was evident. Her
result is easily understood if we take into account that we are only able
 to establish a trend of increasing D/G despite of reaching
much deeper magnitudes.

\begin{figure}
   \centering
   \includegraphics[width=8.5cm]{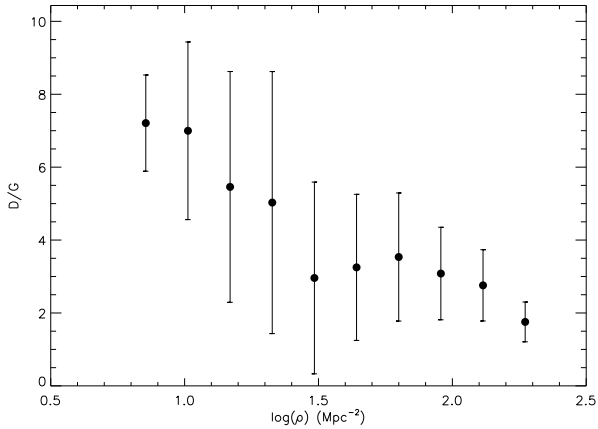}
      \caption{Dwarf to giant ratio as a function of the projected local bright
	galaxy density. The data were binned in $\rho$ to show the relation 
	more clearly.
	}
      \label{dg}
   \end{figure}


\begin{figure}
   \centering
   \includegraphics[width=8.5cm]{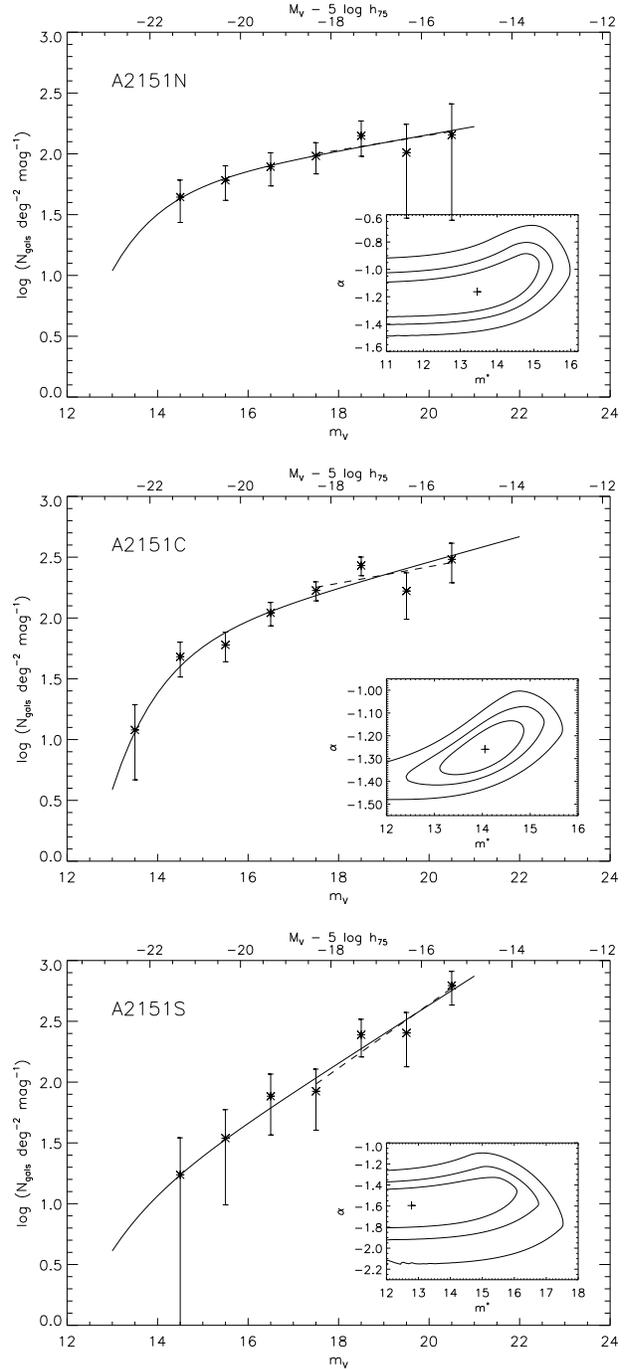}
      \caption{LF of galaxies in each of the subclusters (North to south 
from top to bottom).
	Solid lines are best-fit Schechter functions, while the faint-end slope 
has also been fitted
	to an exponential function (\emph{dashed lines}). The insets show the 1,
 2 and 3$\sigma$ contour
	levels for the best-fitting Schechter parameters.
	}
      \label{A2151_subs_LF}
   \end{figure}


\subsection{D/G ratio as a function of local galaxy density}

Phillips et al.\ (1998) pointed out the existence of a relation
    between the population of dwarf galaxies in a cluster and the
    local density,  analogous to the
    morphology--density relation established for bright galaxies by
    Dressler (1981). This relation implies that dwarf galaxies are
    more common in lower density environments, and is valid for both local
    and medium redshift galaxy clusters (Pracy et al.\ 2004).

We computed the D/G ratio as a function of the projected local
density ($\rho$) for
A2151 (see Fig.~\ref{dg}). The D/G were calculated by computing the 
number
of dwarf galaxies within a circle of radius equal to the distance to
the 10th nearest giant (bright) galaxy. Giant and dwarf galaxies were
defined as in Section 4. The areas for the computation of $\rho$ were
calculated  numerically in order to take the gaps and
overlapping regions in the mosaic into account. Field subtraction was performed
using the counts in the SA107 background field.

It is to be noted from Fig.~\ref{dg} that the projected local
density in A2151 is not as large as in other nearby rich clusters such as
Coma (Trujillo et al.\ 2002). However, our data cover a wide local
density range, reaching the outermost regions of the cluster where
densities are very low. Figure~\ref{dg} also shows that the dwarf
galaxy density relation is only marginal in A2151, with
dwarf galaxies slightly more numerous at lower densities. This
relation is as significant as that derived by Phillips et
al.\ (1998), but not as clean as the one recently derived by Pracy et al.\
(2004)
for A2218.

\subsection{Subclustering}

\begin{figure*}
   \centering
   \includegraphics[width=14cm]{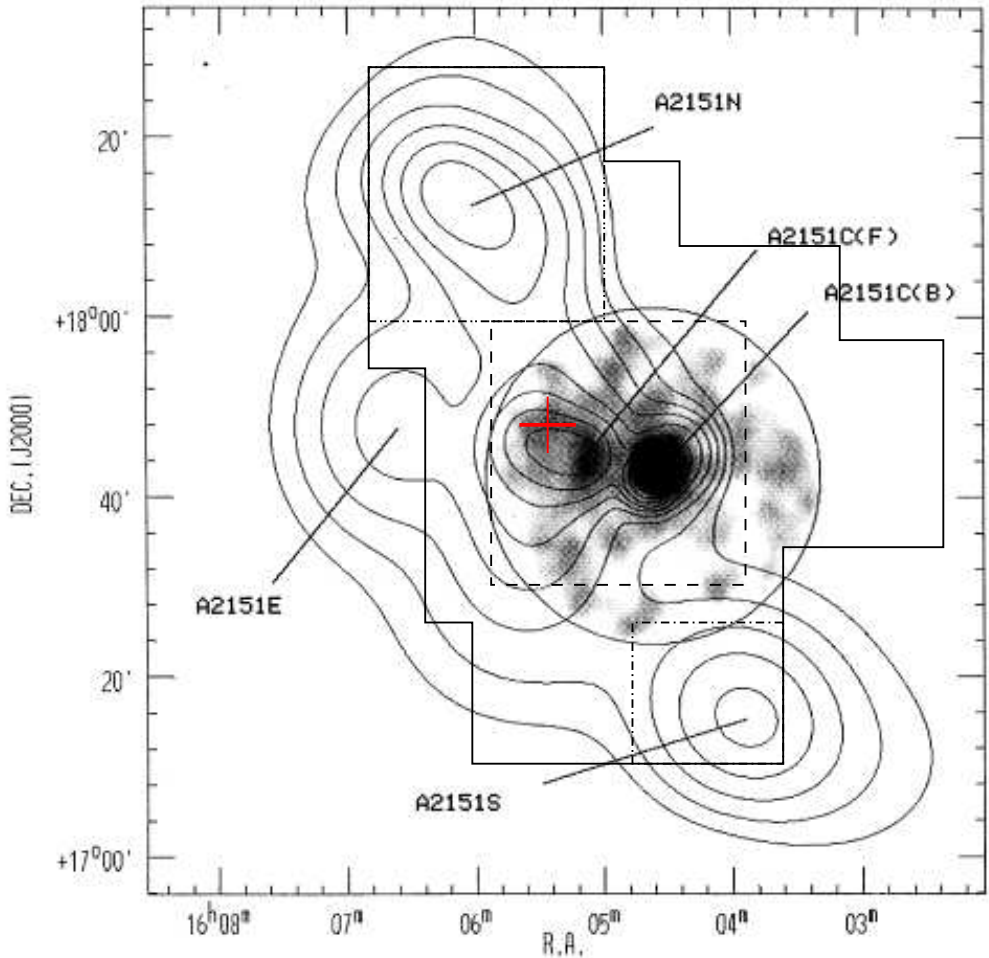}
      \caption{X-ray emission from A2151 (grey-scale distribution) 
      plotted together with galaxy density
	contours (taken from Huang \& Sarazin 1996), showing the significant 
	substructure present in this
	cluster. The LF presented in this paper is computed in the region
	within the solid lines, while the smaller squares (dashed and dotted 
	lines)
	indicate the areas where subclusters' LF were computed.
	}
      \label{Xmap}
   \end{figure*}

In Fig.~\ref{Xmap} we show the X-ray map of the central part of A2151 as 
observed with {\itshape ROSAT} (grey-scale distribution), together with
the luminosity-weighted galaxy density contours (figure taken from Huang 
\& Sarazin 1996). This figure clearly
shows that A2151 exhibits substantial subclustering with at least three 
distinct subclusters identified in the dynamical
analysis of BDB. In our mapping, we cover two of them (A2151N and A2151C,
 see Fig.~\ref{Xmap}) while A2151E is out of our survey.
However, as can be seen from the luminosity-weighted galaxy density 
contours, a fourth overdensity
(namely A2151S) seems to exist in the southernmost region of the cluster,
 though it was not identified as a distinct entity by the
KMM algorithm of BDB.

\begin{figure}
   \centering
   \includegraphics[width=9cm]{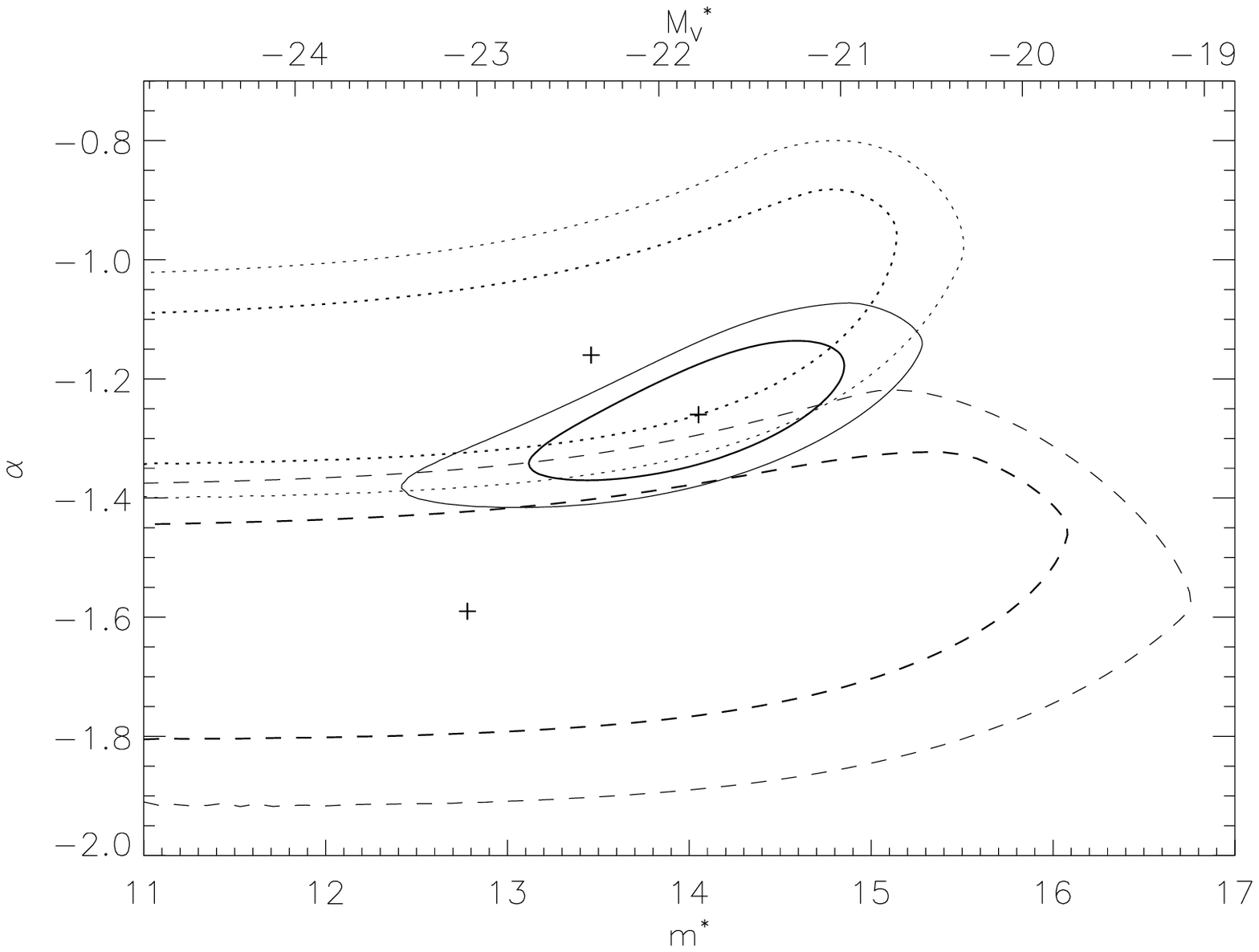}
      \caption{ 68\% and 95\% confidence contours for the best-fit 
      Schechter parameters of A2151N
	(\emph{dotted}), A2151C (\emph{solid}) and A2151S (\emph{dashed}). It 
	can be seen that the characteristic magnitude
	$M^*$ is highly unconstrained in the northern and southern subclusters.
	The crosses mark the best-fit values in each region.
	}
      \label{A2151_contours}
   \end{figure}

As we have shown that the radial LF steepens slightly in the outer 
regions, we computed the LFs of A2151N, A2151C and (partly)
A2151S in the areas delimited by the small squares (dashed and dotted 
lines) in Fig.~\ref{Xmap} in order to look for variations among the 
subclusters.

The results of such analysis are shown in Table~\ref{subparams} and 
in Figure~\ref{A2151_subs_LF}. We find that there is a
tendency towards $\alpha$ increasing from the northern to the southern 
subcluster, although this difference is less significant than
2$\sigma$ (see Fig.~\ref{A2151_contours}). From this figure it is also 
evident that $M^*$ is highly unconstrained in A2151N
and A2151S, which means that an exponential law provides a better 
description of their LFs. With respect to the central subcluster,
its LF presents parameters similar to those of the global LF, revealing 
that A2151C might be its major contributor.

As for the radial LFs, we computed the D/G ratio in the three 
subclusters to confirm that A2151S shows the largest
ratio of all of them (see Table~\ref{subparams}). A more detailed 
inspection of its LF points out that the cause of this enhancement is an 
increase in the
population of faint galaxies in the southernmost subcluster with
respect to the others.


\section{Discussion and conclusions}

We studied the LF of galaxies in A2151 following four different
approaches: the analysis of the global LF, its radial variation, and the 
dependence
with subclustering and local density.

The global LF of galaxies in the cluster exhibits a non-flat faint-end
slope ($\alpha = -1.29$). This is not in line with recent results 
by Valotto et
al.\ (2004), who claim that clusters with significant X-ray emission (that
 of A2151 or greater) present flat
LFs ($\alpha \simeq -1$) while steeper slopes are due to projection
effects in non X-ray emitting clusters. In the literature there are
few measurements of the LF of cluster galaxies in the $V$-band. It is
important to compare results from the same bandpass, because the
parameters of the LF are known to change with wavelength
(Beijersbergen et al.\ 2002). Considering clusters in the nearby
Universe, Lobo et al.\ (1997) study the $V$-band LF in the
central part of Coma, and find a faint-end slope $\alpha_V = -1.59$,
significantly steeper than our value for A2151. However, this result
has been challenged recently by other authors (Beijersbergen et al.\ 2002 ;
 Iglesias-P\'aramo et al.\ 2003; Mobasher et al.\ 2003). At
intermediate redshift, Mercurio et al.\ (2003) analyse the LF of ABCG209
at $z = 0.21$ and find $\alpha_R = -1.20$, which is consistent with
the value presented in this paper.\footnote{It should be noticed that at 
$z \sim 0.2$ the rest-frame $V$-band is redshifted almost to $R$.} In
another recent paper, Pracy et al.\ (2004) compute the LF of the $z =
0.18$ cluster A2218, finding a somewhat steeper slope of $\alpha =
-1.38$, still compatible with ours within the errors. Indeed, comparisons of LFs
of clusters at different redshifts and distinct evolutionary 
states raise the problem of the possible evolution in time of galaxies in the 
cluster, but we simply intend to point out that the values presented in this paper
are similar, within the errors, to those found for other clusters in 
different global conditions.

The radial LF of A2151 shows a trend of increasing faint-end slope as we 
move outwards in the cluster.
 Similar results were obtained in other clusters,
 where the slope increases even more significantly towards the outermost regions 
(Sabatini et al.\ 2003; Andreon 2001; Mercurio et al.\ 2003;
Beijersbergen et al.\ 2002). This has usually been explained 
as dwarf galaxies being destroyed in the inner parts of the clusters,
though it is important to notice that $\alpha$ (or the D/G ratio) can
vary due to changes both in the dwarf or/and giant population.
We have therefore computed the number of dwarf and giant galaxies as a
function of the
projected radial distance to the centre of A2151. The dwarf galaxy 
population in A2151 has a more extended
distribution than giant galaxies, resulting in an increase of the D/G 
ratio as we move outwards.

Alternatively, we investigated the relation between the D/G
    ratio and the projected local giant galaxy density. As expected,
    we  found
    that A2151 has more dwarf galaxies in the lower density
    environments, but this relation is found to be only marginal.

The LFs of the three studied subclusters present dissimilar values of
$\alpha$. Although this difference is only marginally significant (less 
than
2$\sigma$), there is a trend of the faint-end slope to increase from the
northern to the southern subcluster, where faint galaxies are more 
abundant than in the other two subclusters.

We plan to extend this study in the future to the rest of the Hercules 
Supercluster, where the environment
should clearly reveal its influence. The combination of this data with 
morphologies and velocities for cluster members will help us
to shed light on the evolutionary status of galaxies within the 
Supercluster.


\begin{acknowledgements}
This work has made use of the NASA/IPAC Extragalactic Database (NED) 
which is operated by the Jet Propulsion Laboratory,
California Institute of Technology, under contract with the National 
Aeronautics and Space Administration. The INT is operated
on the island of La Palma by the ING group, in the Spanish Observatorio 
del Roque de Los Muchachos of the Instituto de
Astrof\'\i sica de Canarias. This work has made use of the ING data 
archive in Cambridge and of the ING service programme at ORM.
This paper has been funded by the Spanish DGES, grant AYA2001-3939. RSJ 
acknowledges support from the Academia Canaria de Ciencias.
\end{acknowledgements}

\newpage

\onecolumn




\clearpage

\end{document}